\documentclass[conference]{IEEEtran}

\usepackage[switch,columnwise]{lineno}

\usepackage{color}
\usepackage{comment}
\usepackage{cite}
\usepackage{multirow}
\usepackage{graphicx}
\usepackage{caption}
\usepackage{subcaption}
\usepackage{epstopdf}
\usepackage{verbatim}
\usepackage{amsmath}
\usepackage{hyphenat}
\usepackage{fancyvrb}
\usepackage{boxedminipage}
\usepackage{xcolor}
\usepackage{amsmath}
\usepackage{algorithm}
\usepackage[noend]{algpseudocode}

\usepackage{array}
\usepackage{amssymb}

\makeatletter
\def\endthebibliography{%
	\def\@noitemerr{\@latex@warning{Empty `thebibliography' environment}}%
	\endlist
}
\makeatother

\begin{document}

\title{Large-Scale Discrete Fourier Transform on TPUs}

\author{Tianjian~Lu,~\IEEEmembership{}
Yi-Fan~Chen,~\IEEEmembership{}
Blake~Hechtman,~\IEEEmembership{}
Tao~Wang,~\IEEEmembership{}
and~John~Anderson~\IEEEmembership{}
\thanks{The authors are with Google Research, 1600 Amphitheatre Pkwy, Mountain View, CA 94043,  USA, e-mail: \{tianjianlu, yifanchen, blakehechtman, wangtao, janders\}@google.com.}
}

\author{
	\IEEEauthorblockN{Tianjian~Lu,~Yi-Fan~Chen,~Blake~Hechtman,~Tao~Wang,~and~John~Anderson}
	\IEEEauthorblockA{Google Research
		\\1600 Amphitheatre Pkwy, Mountain View, CA 94043, USA
		\\e-mail: \{tianjianlu, yifanchen, blakehechtman, wangtao, janders\}@google.com}
}

\markboth{}%
{Shell \MakeLowercase{\textit{Lu et al.}}: Large-Scale Discrete Fourier Transform on TPUs}

\maketitle

\thispagestyle{plain}
\pagestyle{plain}

\begin{abstract}
In this work, we present two parallel algorithms for the large-scale discrete Fourier transform (DFT) on Tensor Processing Unit (TPU) clusters.
The two parallel algorithms are associated with two formulations of DFT:
one is based on the Kronecker product, to be specific, dense matrix multiplications between the input data and the Vandermonde matrix, denoted as KDFT in this work;
the other is based on the famous Cooley-Tukey algorithm and phase adjustment, denoted as  FFT in this work.
Both KDFT and FFT formulations take full advantage of TPU's strength in matrix multiplications.
The KDFT formulation allows direct use of nonuniform inputs without additional step.
In the two parallel algorithms, the same strategy of data decomposition is applied to the
input data.
Through the data decomposition, the dense matrix multiplications in KDFT and FFT are kept local within TPU cores, which can be performed completely in parallel.
The communication among TPU cores is achieved through the one-shuffle scheme in both parallel algorithms,
with which sending and receiving data takes place simultaneously between two neighboring cores and along the same direction on the interconnect network.
The one-shuffle scheme is designed for the interconnect topology of TPU clusters, minimizing the time required by the  communication among TPU cores.
Both KDFT and FFT are implemented in TensorFlow.
The three-dimensional complex DFT is performed on an example of dimension $8192 \times 8192 \times 8192$  with a full TPU Pod:
the run time of KDFT is 12.66 seconds and that of FFT is 8.3 seconds.
Scaling analysis is provided to demonstrate the high parallel efficiency of the two DFT implementations on TPUs.
\end{abstract}



\section{Introduction}
The discrete Fourier transform (DFT) is critical in many scientific and engineering applications, including time series and waveform analyses,
convolution and correlation computations,  solutions to partial differential equations, density function theory in first-principle calculations,
spectrum analyzer, synthetic aperture radar, computed tomography, magnetic resonance imaging,
and derivatives pricing \cite{bracewell1986fourier,  grama2003introduction, takahashifast, cont2009frontiers}.
However, the computation efficiency of DFT is often the formidable bottleneck in handling large-scale problems due to the large data size
and real-time-processing requirement \cite{giannakis2014signal, olshannikova2015visualizing}.
In general,  efforts on enhancing the computation efficiency of  DFT fall into two categories:
seeking fast algorithms and adapting the  fast algorithms to hardware accelerators.
One breakthrough of  the fast-algorithm category is the Cooley-Tukey algorithm \cite{cooley1965algorithm}, also known as the fast Fourier transform (FFT),
which reduces the complexity of $N$-point DFT from $O(N^2)$ to $O(N\log N)$.
The Cooley-Tukey algorithm assuming that the number of data is a power of two is known as the Radix-2 algorithm 
and followed by Mixed-Radix  \cite{takahashifast} and Split-Radix \cite{duhamel1984split} algorithms.

In addition to the fast algorithms, the performance of hardware accelerators has been steadily driving the efficiency enhancement of DFT computation:
the first implementation of the FFT algorithm was realized on ILLIAC IV parallel computer \cite{ackins1968fast, stevens1971fast};
over the years, the DFT computation has been adapted to both shared-memory \cite{swarztrauber1987multiprocessor, bailey1990ffts} and
distributed-memory architectures \cite{gupta1993scalability, frigo2005design, pekurovsky2012p3dfft, takahashi2009implementation, popovici2019flexible}.
The advancement of hardware accelerators has enabled massive parallelization for DFT computation.
One such example is deploying the FFT computation on manycore processors \cite{jeongnimkim2018leveraging}.
Another example is implementing the FFT algorithm on clusters of graphics processing units (GPUs) \cite{roe2019multigpu}.
A GPU cluster contains a number of nodes (machines) and within each node, GPUs are connected through PCIe, a high-speed serial interface.
The Cooley-Tukey algorithm and its variants often require a large number of memory accesses per arithmetic operation such that
the bandwidth limitation of PCIe becomes the computation bottleneck of the overall performance of FFT on GPU clusters.
Prior to the recent development of novel high-speed interconnects such as NVLink \cite{foley2017ultra, li2019evaluating},
many efforts related to the GPU-accelerated  DFT computation are spent on minimizing the PCIe transfer time \cite{takahashifast, chen2010large}.
It is worth mentioning that the route of algorithm-hardware co-design has also been taken with Field Programmable Gate Arrays (FPGAs) to optimize the configurations
of a customized hardware accelerator for high-performance computing of DFT \cite{nag2000method, garrido2014challenging, yu2010multidimensional}.

The recent success of machine learning (ML), or deep learning (DL) in particular, has spurred a new wave of hardware accelerators.
In many ML applications, it becomes increasingly challenging to balance the performance-cost-energy of processors with the growth of data.
Domain-specific hardware is considered as a promising approach to achieve this \cite{stoica2017berkeley}.
One example of the domain-specific hardware is Google's Tensor Processing Unit (TPU) \cite{jouppi2017datacenter}.
As a reference, TPU v3 provides 420 teraflops and 128 GiB high-bandwidth memory (HBM) \cite{tpuv3}.
In witnessing how DFT computation benefits from the development of hardware accelerators,
it is tempting to ask whether TPU can empower the large-scale DFT computation.
It is plausible with the following four reasons:
(1) TPU is an ML application-specific integrated circuit (ASIC), devised for neural networks (NNs);
NNs require massive amounts of multiplications and additions between the data and parameters and 
TPU can handle these computations in terms of matrix multiplications in a very efficient manner \cite{jouppi2017quantifying};
similarly, DFT can also be formulated as matrix multiplications between the input data and the Vandemonde matrix;
(2) TPU chips are connected directly to each other with dedicated, high-speed, and low-latency interconnects, bypassing host CPU or any networking resources;
therefore, the large-scale DFT computation can be distributed among multiple TPUs with minimal communication time and hence very high parallel efficiency;
(3) the large capacity of the in-package memory of TPU makes it possible to handle large-scale DFT efficiently;
and (4) TPU is programmable with software front ends such as TensorFlow \cite{wu2016google} and PyTorch \cite{ketkar2017introduction}, both of which make it straightforward
to implement the parallel algorithms of DFT on TPUs.
In fact, all the aforementioned four reasons have been verified in the high-performance
Monte Carlo simulations on TPUs \cite{yang2019high, belletti2019tensor}.

In this work, we present two parallel algorithms for DFT on TPUs:
one is based on the Kronecker product, to be specific, dense matrix multiplications between the input data and the Vandermonde matrix, denoted as KDFT in this work;
and the other is based on the  Cooley-Tukey algorithm and phase adjustment, denoted as FFT in this work.
For a $N$-point DFT, the computation complexity of KDFT is $O(N^2)$, whereas that of FFT is $O(N \log N)$.
Both parallel algorithms take full advantage of TPU's strength in matrix multiplications.
It is worth mentioning that KDFT takes in nonuniform input data without additional steps.
The nonuniform Fourier transform has important applications in signal processing, medical imaging, numerical solutions of partial differential equations, and machine learning \cite{bagchi1996nonuniform1, bagchi1996nonuniform2, lee2005type, rahimi2008random}.
Both KDFT and FFT use the same strategy of data decomposition over the input data,
through which the dense matrix multiplications are kept local within TPU cores and can be performed completely in parallel.
The communication among TPU cores is achieved through the one-shuffle scheme in both parallel algorithms,
with which sending and receiving data takes place simultaneously between two neighboring cores and along the same direction on the interconnect network.
The one-shuffle scheme is designed for the interconnect topology of TPU clusters, minimizing the time required by the  communication among TPU cores.
Scaling analysis is provided to demonstrate the high parallel efficiency of the proposed two algorithms of DFT on TPUs.

\section{TPU System Architecture}

\begin{figure}[t!]
	\centering
	\begin{subfigure}{\linewidth}
		\centering
		\includegraphics[width=0.9\linewidth]{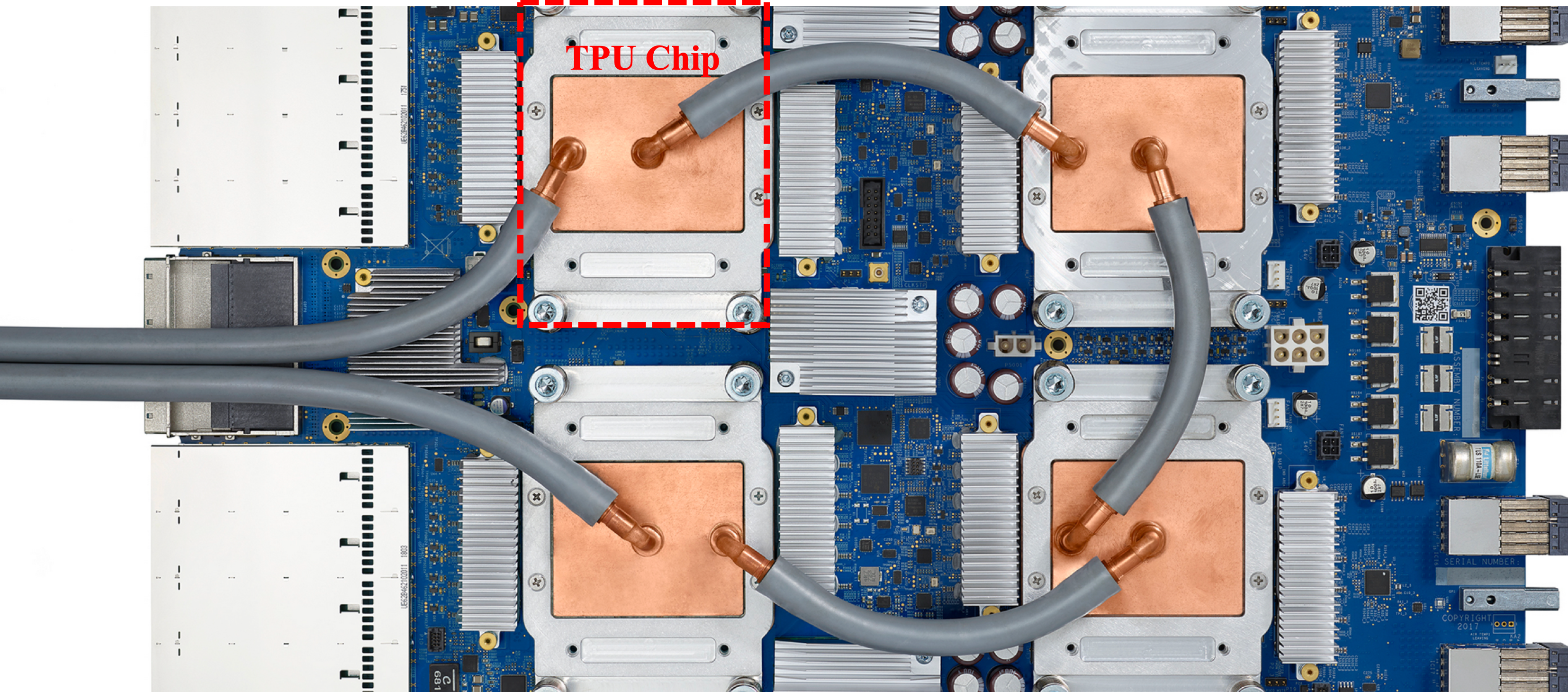}
		\caption{} 
	\end{subfigure}\vfill
	\begin{subfigure}{\linewidth}
		\centering
		\includegraphics[width=0.9\linewidth]{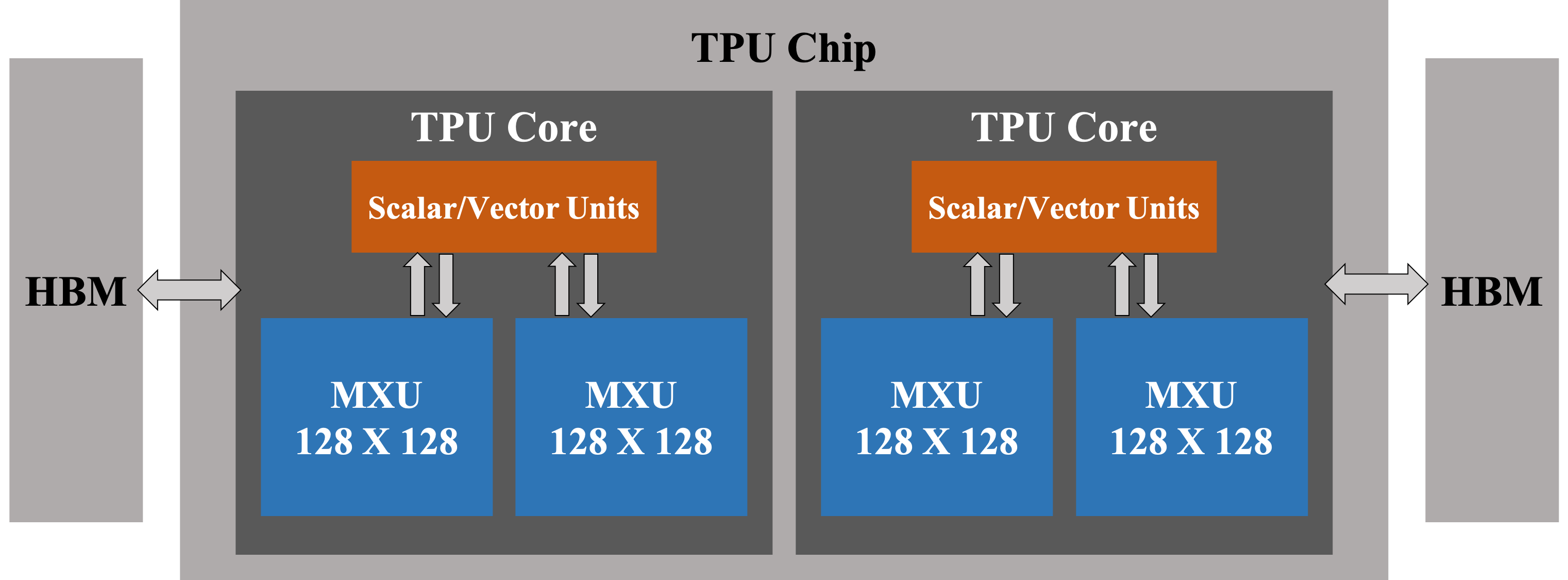}
		\caption{} 
	\end{subfigure}
	\caption{(a) TPU v3 has four chips on the same board and (b) each chip contains two cores.}
	\label{tpu_board_chip}
\end{figure}

In this section, we provide an overview of the TPU system architecture on both the hardware and software components.

\subsection{Hardware architecture}
Figure \ref{tpu_board_chip} shows one TPU board or unit:
there are four TPU chips on the same board; each chip has two cores; and each core contains the scalar, vector, and matrix units (MXU).
Structured as a $128 \times 128$ systolic array, MXU provides the bulk of compute power of a TPU chip and handles 16 K multiply-accumulate (MAC) operations in one clock cycle.
The inputs and outputs of MXU are float32 and the MAC operations on MXU are performed with bfloat16 \cite{bfloat16}.
However, one float32 number can be decomposed into multiple bfloat16 numbers and with appropriate accumulations, high-precision MAC operation can be achieved \cite{henry2019leveraging}.
The implementation of both parallel algorithms in this work leverages the strategy of decomposition and accumulation and achieves the precision of float32.
As shown in Fig.\ \ref{tpu_board_chip}(b), each TPU core has 16 GiB high-bandwidth memory (HBM).
The large capacity of in-package memory makes it possible to solve large-scale problems in a highly efficient manner.
TPU is designed as a coprocessor on the I/O bus: each board shown in Fig. \ref{tpu_board_chip}(a) is paired with one host server
consisting of CPU, RAM, and hard disk; TPU executes the instructions sent from CPU on the host server through PCIe.
 
\begin{figure}[t!]
	\includegraphics[width=8cm, height=3cm]{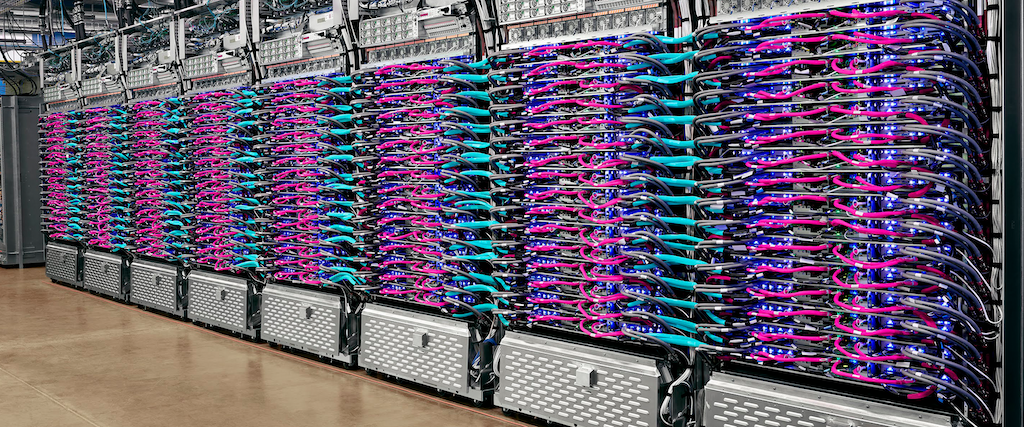}
	\centering
	\caption{TPU v3 Pod in a data center.}
	\label{tpu_pod}
\end{figure}

Figure \ref{tpu_pod} shows a TPU v3 Pod in a data center where a total number of 2048 cores are connected to each other.
In a Pod configuration, TPU chips are connected through dedicated high-speed interconnects of very low latency.
The interconnect topology is a two-dimensional (2D) toroidal mesh with each chip connected to its four nearest neighbors such that the communication takes place in four directions.
These interconnects bypass the CPU networking resources and go from chip to chip directly.
In our implementations, we have further optimized the communication strategy to take advantage of the TPU interconnect topology.

\begin{figure}[t!]
	\includegraphics[width=8cm, height=6cm]{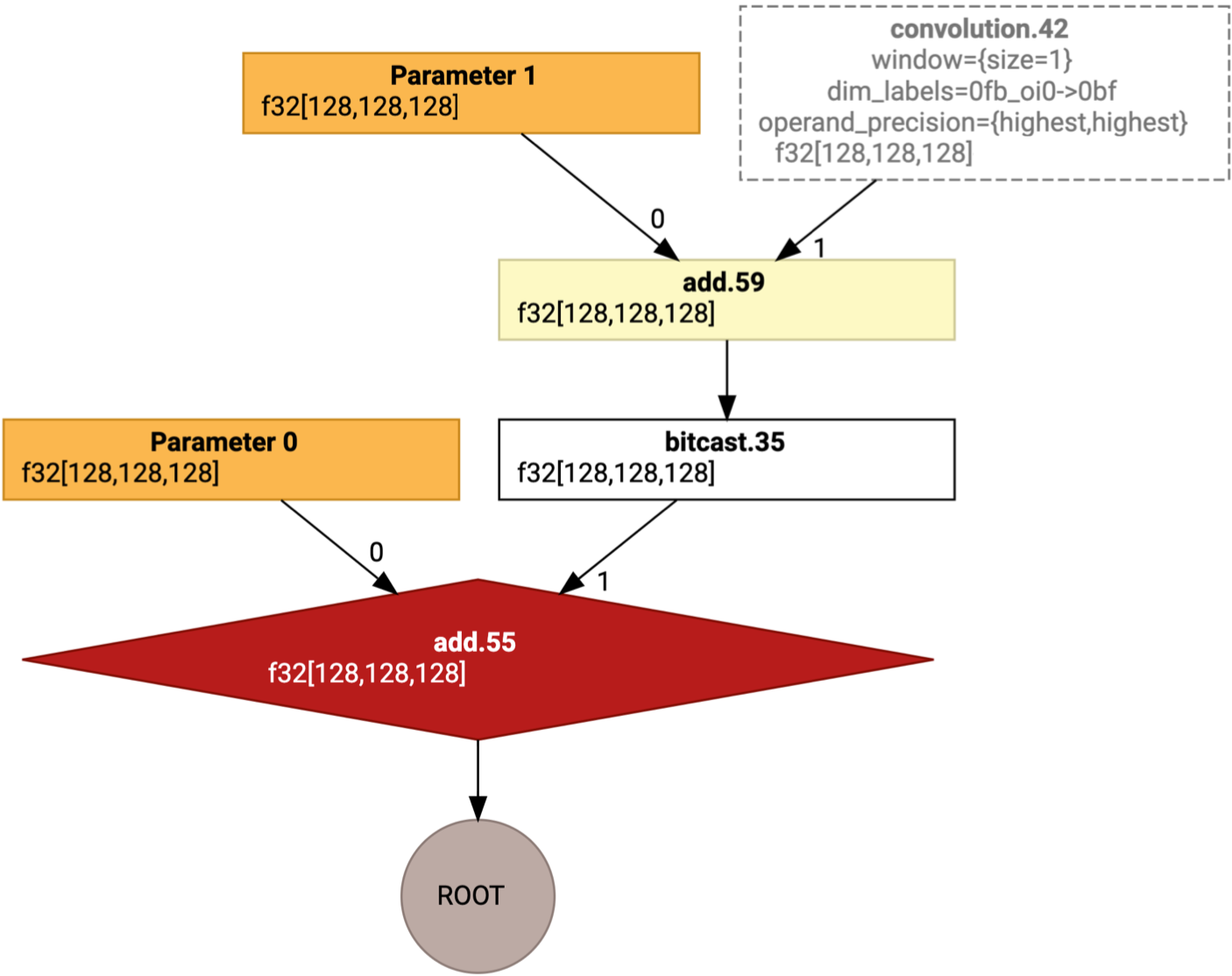}
	\centering
	\caption{A computation graph of TensorFlow operations.}
	\label{sample_graph}
\end{figure}

\subsection{Software architecture}
TensorFlow is used to program TPUs in this work.
A TensorFlow client converts the TensorFlow operations into a computational graph.
A sample computation graph performing addition and convolution operations is shown in Fig.\ \ref{sample_graph}.
The TensorFlow client sends the  graph to a TensorFlow server.
The TensorFlow server partitions the computational graph into portions that run on TPU and CPU, respectively.
If multiple TPUs are to be employed, the graph is marked for replication.
The TensorFlow server then compiles the sub-graph that runs on TPUs into a HLO program
and  invokes the accelerated linear algebra (XLA) compiler.
The XLA compiler takes in the HLO program and converts it into a LLO program, which is effectively the assembly code for TPUs.
Both the generation and compilation of the computational graph occur on the host server.
The compiled LLO code is loaded onto TPUs for execution from the host server through PCIe.

The memory usage of a TPU is determined at compile time.
Because both the hardware structure of MXU and the memory subsystem on a TPU core prefer certain shapes of a tensor variable involved in an operation,
the XLA compiler performs the data layout transformations in order for the hardware to efficiently process the operation.
If a tensor variable does not align with the preferred shape, the XLA compiler pads zeros along one dimension to make it a multiple of eight and the other dimension to a multiple of 128.
Zero padding under-utilizes the TPU core and leads to sub-optimal performance, which should be taken into account in the implementation of the parallel algorithms on TPUs.

\section{DFT Formulations}
In this section, we provide the detailed formulations for both KDFT and FFT.

\subsection{KDFT Formulation}
The KDFT formulation starts from the general form of DFT, which is defined as
\begin{equation}
X_k \triangleq X(z_{k}) = \sum_{n=0}^{N-1} x_n z_{k}^{-n},
\label{dft1d}
\end{equation}
where $\triangleq$ denotes ``defined to be", $x_n$ represents the input,  and $\left\{z_{k}\right\}_{k=0}^{N-1}$ are $N$ distinctly and arbitrarily sampled points on the $z$-plane.
Equation (\ref{dft1d}) can be rewritten into the matrix form
\begin{equation}
\left\{X\right\} = \left[ V \right] \left\{ x \right\},
\label{dft1d_matrix}
\end{equation}
where
\begin{eqnarray}
\left\{X\right\}  \!\!\!&=&\!\!\! \left( X(z_0), X(z_1), \cdots, X(z_{N-1}) \right)^{\textrm{T}}, \nonumber \\
\left\{x\right\}  \!\!\!&=\!\!\!& \left( x_0, x_1, \cdots, x_{N-1} \right)^{\textrm{T}}, \nonumber
\end{eqnarray}
and
\begin{equation}
\left[ V \right] = 
\begin{pmatrix}
1 & z_0^{-1} & z_0^{-2} & \cdots & z_0^{-(N-1)} \\
1 & z_1^{-1} & z_1^{-2} & \cdots & z_1^{-(N-1)} \\
\vdots & \vdots & \vdots & \ddots & \vdots \\
1 & z_{N-1}^{-1} & z_{N-1}^{-2} & \cdots & z_{N-1}^{-(N-1)} \\
\end{pmatrix}.
\end{equation}
Note that $\left[ V \right]$ is the Vandermonde matrix of dimension $N \times N$.
Computing the inverse DFT is equivalent to solving the linear system in Equation (\ref{dft1d_matrix}).
In the situation when $\left\{z_{k}\right\}_{k=0}^{N-1}$ are uniformly sampled on the $z$-plane,  the Vandermonde matrix $\left[ V \right]$ becomes unitary and contains the roots of unity.

The general form of a 2D DFT can be written as
\begin{equation}
X(z_{1k}, z_{2k}) = \sum_{n_1=0}^{N_1-1} \sum_{n_2=0}^{N_2-1} x(n_1, n_2) z_{1k}^{-n_1} z_{2k}^{-n_2},
\label{dft2d}
\end{equation}
where $\left[x\right]$ has the dimension of $N_1 \times N_2$ and
$\left\{\left( z_{1k}, z_{2k}\right)\right\}_{k=0}^{N_1N_2-1}$ represents the set of distinctly and arbitrarily sampled points in $\left( z_1, z_2 \right)$ space.
It is worth mentioning that the sampling with $\left( z_{1k}, z_{2k}\right)$ has to ensure the existence of the inverse DFT. If the sampling is performed on rectangular grids, Equation (\ref{dft2d}) can be rewritten into the matrix form as
\begin{equation}
\left[ X \right] = \left[ V_1 \right] \left[ x \right]  \left[ V_2 \right]^{\textrm{T}},
\label{dft2d_matrix}
\end{equation}
where
\begin{equation}
\left[ X \right] = \left(
\begin{smallmatrix}
X(z_{10}, z_{20}) & X(z_{10}, z_{21}) & \cdots & X(z_{10}, z_{2, N_2-1}) \\
X(z_{11}, z_{20}) & X(z_{11}, z_{21})  & \cdots & X(z_{11}, z_{2, N_2-1}) \\
\vdots & \vdots  & \ddots & \vdots \\
X(z_{1, N_1-1}, z_{20}) & X(z_{1, N_1-1}, z_{21}) & \cdots & X(z_{1, N_1-1}, z_{2, N_2-1}) \\
\end{smallmatrix} \right),
\end{equation}
\begin{equation}
\left[ x \right] = \left(
\begin{smallmatrix}
x(0, 0) & x(0, 1) & \cdots & x(0, N_2-1) \\
x(1, 0) & x(1, 1) & \cdots & x(1, N_2-1) \\
\vdots & \vdots  & \ddots & \vdots \\
x(N_1-1, 0) & x(N_1-1, 1) & \cdots & x(N_1-1, N_2-1) \\
\end{smallmatrix} \right),
\end{equation}
\begin{equation}
\left[ V_1 \right] = 
\begin{pmatrix}
1 & z_{10}^{-1} & z_{10}^{-2} & \cdots & z_{10}^{-(N_1-1)} \\
1 & z_{11}^{-1} & z_{11}^{-2} & \cdots & z_{11}^{-(N_1-1)} \\
\vdots & \vdots & \vdots & \ddots & \vdots \\
1 & z_{1, N_1-1}^{-1} & z_{1, N_1-1}^{-2} & \cdots & z_{1, N_1-1}^{-(N_1-1)} \\
\end{pmatrix},
\end{equation}
and
\begin{equation}
\left[ V_2 \right] = 
\begin{pmatrix}
1 & z_{20}^{-1} & z_{20}^{-2} & \cdots & z_{20}^{-(N_2-1)} \\
1 & z_{21}^{-1} & z_{21}^{-2} & \cdots & z_{21}^{-(N_2-1)} \\
\vdots & \vdots & \vdots & \ddots & \vdots \\
1 & z_{2, N_2-1}^{-1} & z_{2, N_2-1}^{-2} & \cdots & z_{2, N_2-1}^{-(N_2-1)} \\
\end{pmatrix}.
\end{equation}
Note that both $\left[ V_1 \right]$ and $\left[ V_2 \right]$ are Vandermonde matrices of dimensions $N_1 \times N_1$ and $N_2 \times N_2$, respectively.
One can further lump $\left[x\right]$ into a vector such that Equation (\ref{dft2d_matrix}) can be written into the same matrix form as Equation (\ref{dft1d_matrix}),
in which $\left[ V \right]$ for the 2D DFT is expressed as
\begin{equation}
\left[ V \right] = \left[ V_1 \right] \otimes \left[ V_2 \right],
\end{equation}
where $\otimes$ denotes the Kronecker product \cite{regalia1989kronecker}.

Similarly, the three-dimensional (3D) DFT defined by
\begin{equation}
X(z_{1k}, z_{2k}, z_{3k}) = \sum_{n_1=0}^{N_1-1} \sum_{n_2=0}^{N_2-1} \sum_{n_3=0}^{N_3-1} x(n_1, n_2, n_3) z_{1k}^{-n_1} z_{2k}^{-n_2} z_{3k}^{-n_3}
\label{dft3d}
\end{equation}
can be rewritten into the matrix form as
\begin{equation}
\left\{X\right\} = \left[ V_1 \right] \otimes \left[ V_2 \right] \otimes \left[ V_3 \right] \left\{ x \right\},
\label{dft3d_matrix}
\end{equation}
where
\begin{eqnarray}
\left[ V_j \right] \!\!\!&=&\!\!\! 
\begin{pmatrix}
1 & z_{j0}^{-1} & z_{j0}^{-2} & \cdots & z_{j0}^{-(N_j-1)} \\
1 & z_{j1}^{-1} & z_{j1}^{-2} & \cdots & z_{j1}^{-(N_j-1)} \\
\vdots & \vdots & \vdots & \ddots & \vdots \\
1 & z_{j, N_j-1}^{-1} & z_{j, N_j-1}^{-2} & \cdots & z_{j, N_j-1}^{-(N_j-1)} \\
\end{pmatrix}, \nonumber \\
 j \in \!\!\!\!&&\!\!\!\!\!\!\!\!\!\!\!\! \left\{1, 2, 3\right\}.
\end{eqnarray}
For the 3D DFT defined in Equation (\ref{dft3d_matrix}), the sampling is performed on rectangular grids in $\left( z_1, z_2, z_3 \right)$ space
and the Vandermonde matrix $\left[ V_3 \right]$ has the dimension of $N_3 \times N_3$.
It can be seen that the Kronecker product bridges the gap between the matrix and tensor operations,
through which the contraction between a rank-2 tensor and another rank-3 tensor in the 3D DFT can be formulated as matrix multiplications.
The KDFT formulation can be easily extended to higher dimensions.

\subsection{FFT Formulation}
The FFT formulation starts with
\begin{equation}
X_k \triangleq \sum_{n=0}^{N-1} x_n  e^{\displaystyle -j 2 \pi\frac{n k}{N}},
\label{fft1d}
\end{equation}
in which $x_n$ represents the input data and the frequency sampling has to be uniform.
The global index $n$ in Equation (\ref{fft1d}) can be expressed as
\begin{equation}
n = P l + \beta,
\label{global_index}
\end{equation}
where $l = 0, 1,  \cdots, \frac{N}{P} -1$ and $\beta = 0, 1, \cdots, P-1 $.
With Equation (\ref{global_index}), Equation (\ref{fft1d}) can be rewritten as 
\begin{align}
X_k &\triangleq \sum_{n=0}^{N-1} x_{(P l + \beta)} e^{\displaystyle -j 2 \pi \frac{(Pl + \beta)k}{N}}  \\
&= \sum_{\beta=0}^{P-1} e^{\displaystyle  -j 2 \pi  \frac{\beta k}{N}} \left( \sum_{l=0}^{\frac{N}{P} -1}  x_{(P l + \beta)} e^{-j 2 \pi \displaystyle \frac{l k}{\frac{N}{P}}}  \right).
\label{fft1d_decomposed}
\end{align}
In Equation (\ref{fft1d_decomposed}),
\begin{equation}
\widetilde{X}_k = \sum_{l=0}^{\frac{N}{P} -1}  x_{(P l + \beta)} e^{-j 2 \pi \displaystyle \frac{l k}{\frac{N}{P}}}  
\end{equation}
is computed with the famous Cooley-Tukey algorithm locally on individual cores.
Prior to the local transform,  the gathering of the input among the cores is required,
which arises from the global indexing in Equation (\ref{global_index}).
After the local transform,  the phase adjustment needs to applied, which is formulated as matrix multiplications similar to that in Equation (\ref{dft1d_matrix}).
Higher dimensional FFT such as 2D and 3D can be achieved by repeating this one-dimensional (1D) scheme along the corresponding dimensions.

\begin{figure*}[ht]
	\centering
	\begin{subfigure}[b]{0.5\linewidth}
		\centering
		\includegraphics[width=9cm, height=8cm]{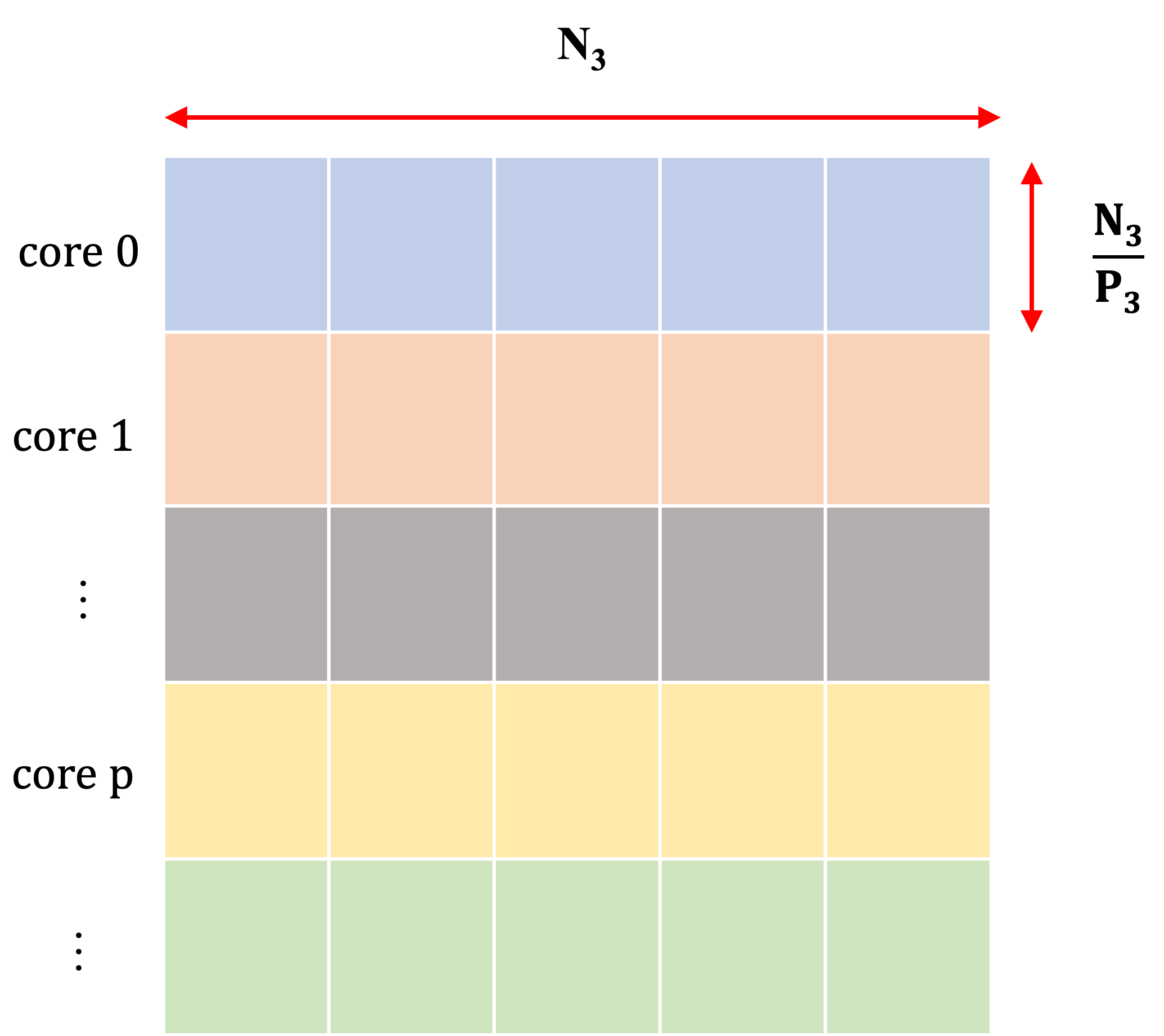}
		\caption{} 
	\end{subfigure}\hfill
	\begin{subfigure}[b]{0.5\linewidth}
		\centering
		\includegraphics[width=7cm, height=6.5cm]{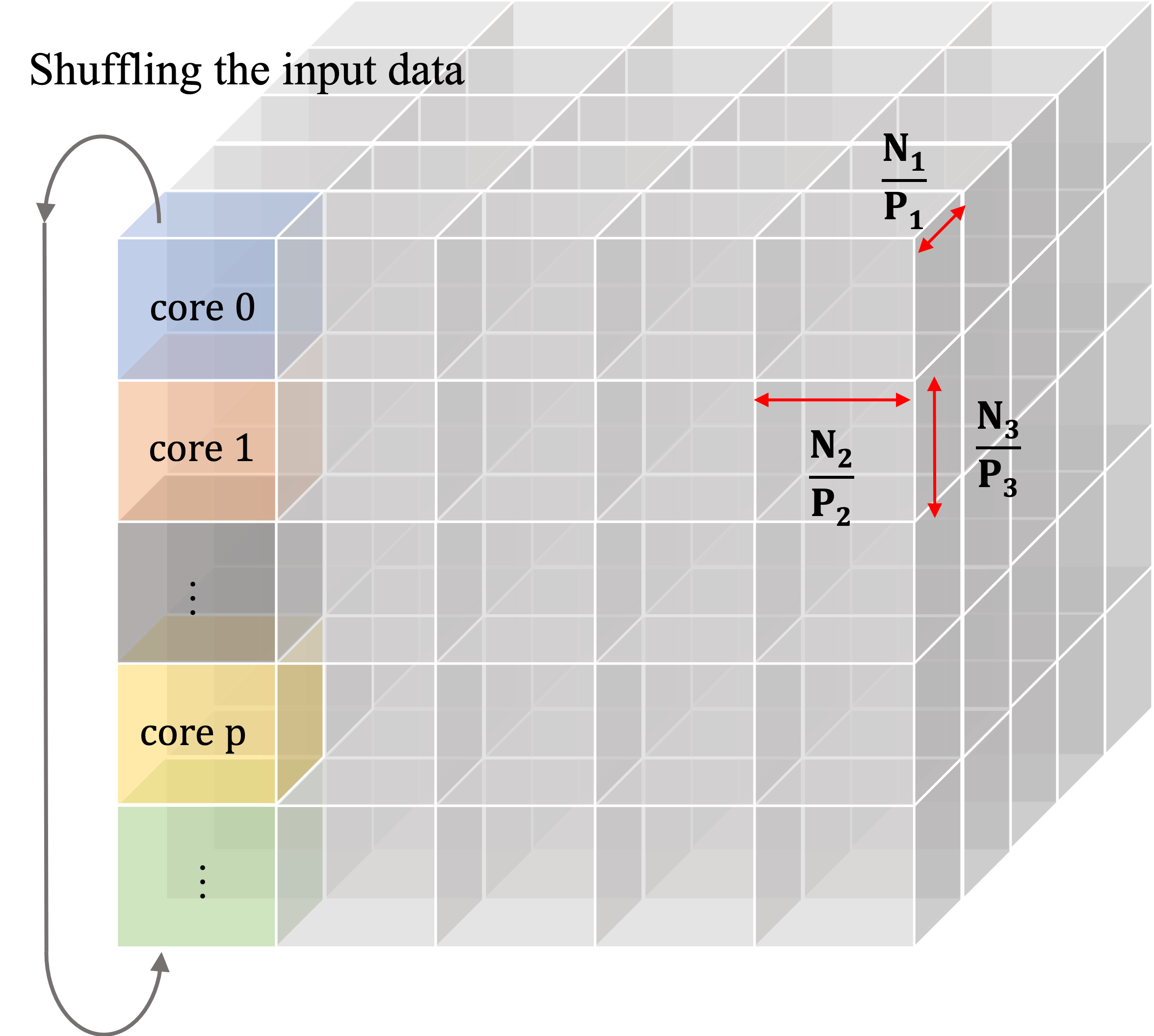}
		\caption{} 
	\end{subfigure}
	\caption{Through the data decomposition with the TPU computation shape $\left( P_1, P_2, P_3 \right)$, each TPU core contains (a) the Vandermonde matrix of dimension $\frac{N_i}{P_i} \times N_i$, $i = 1, 2, 3$ and (b) the block of input data of dimension $\frac{N_1}{P_1} \times \frac{N_2}{P_2}  \times \frac{N_3}{P_3}$ for a 3D DFT.  The core index is denoted by $p = 0, 1, \cdots, P_i-1$, $i = 1, 2, 3$. The Fourier transform along the third dimension requires shuffling the blocks of the input data among the cores that are grouped by the third dimension of the computation shape $P_3$.}
	\label{load_balancing}
\end{figure*}

\section{Implementation of the parallel algorithms}

\begin{figure}[]
	\centering
	\begin{subfigure}{\linewidth}
		\centering
		\includegraphics[width=\linewidth]{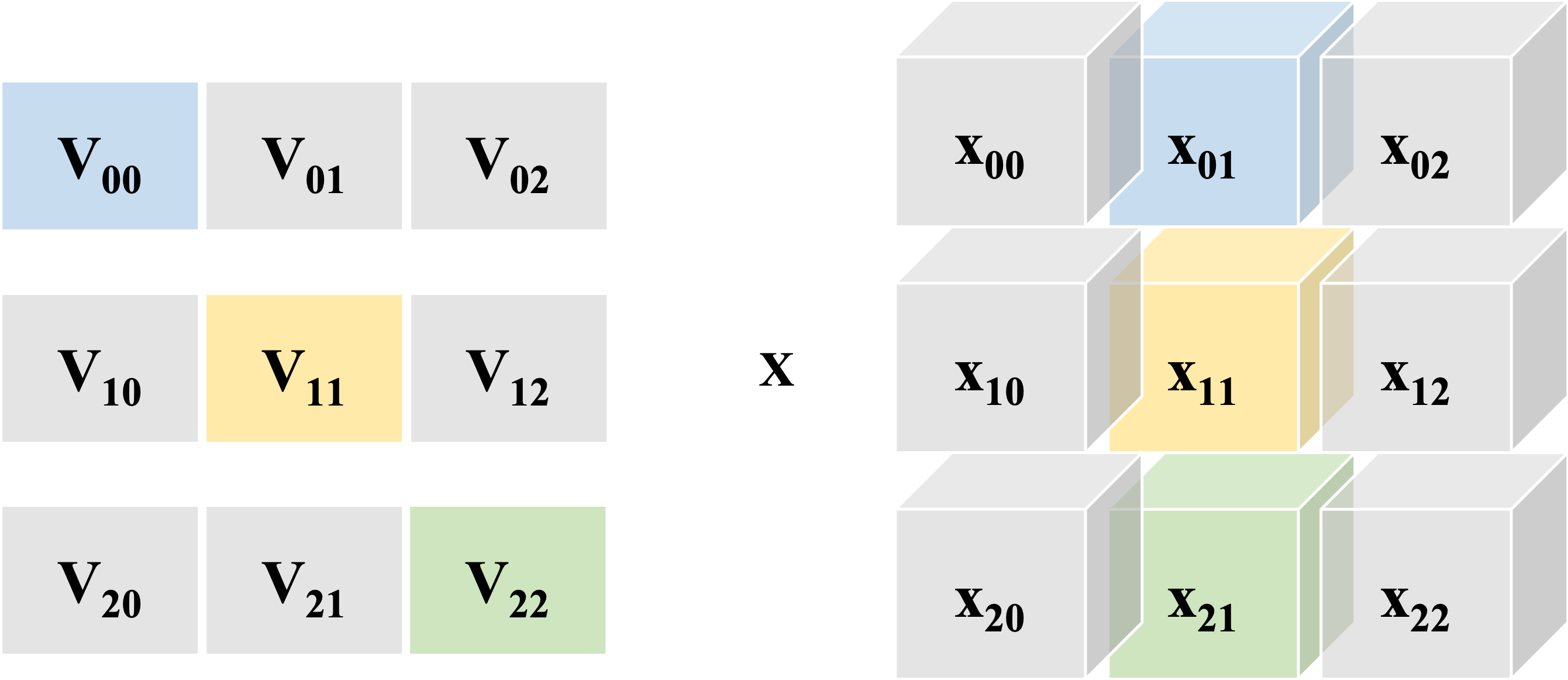}
		\caption{} 
	\end{subfigure}\hfill
	\begin{subfigure}{\linewidth}
		\centering
		\includegraphics[width=\linewidth]{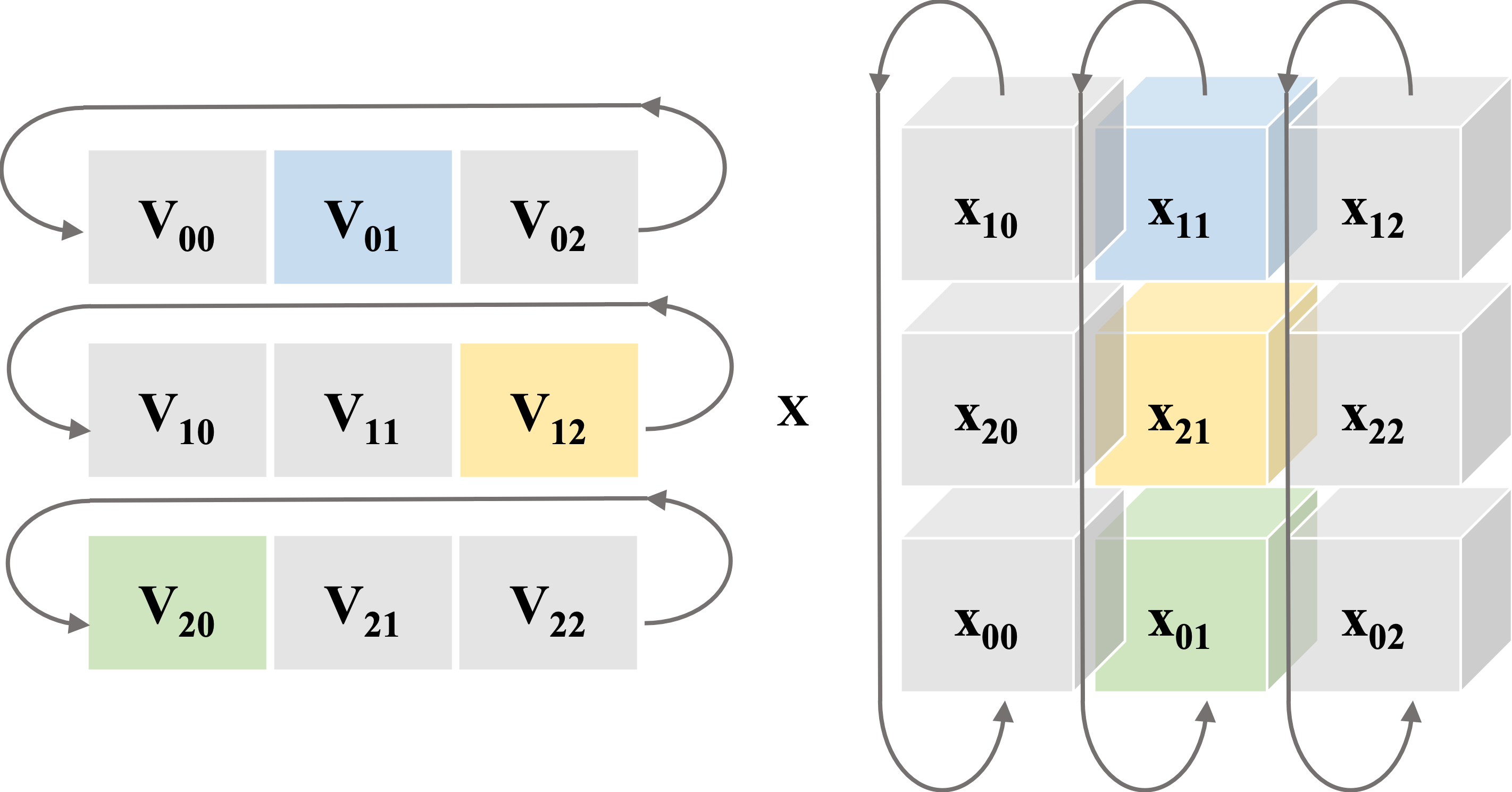}
		\caption{} 
	\end{subfigure}
	\begin{subfigure}{\linewidth}
		\centering
		\includegraphics[width=\linewidth]{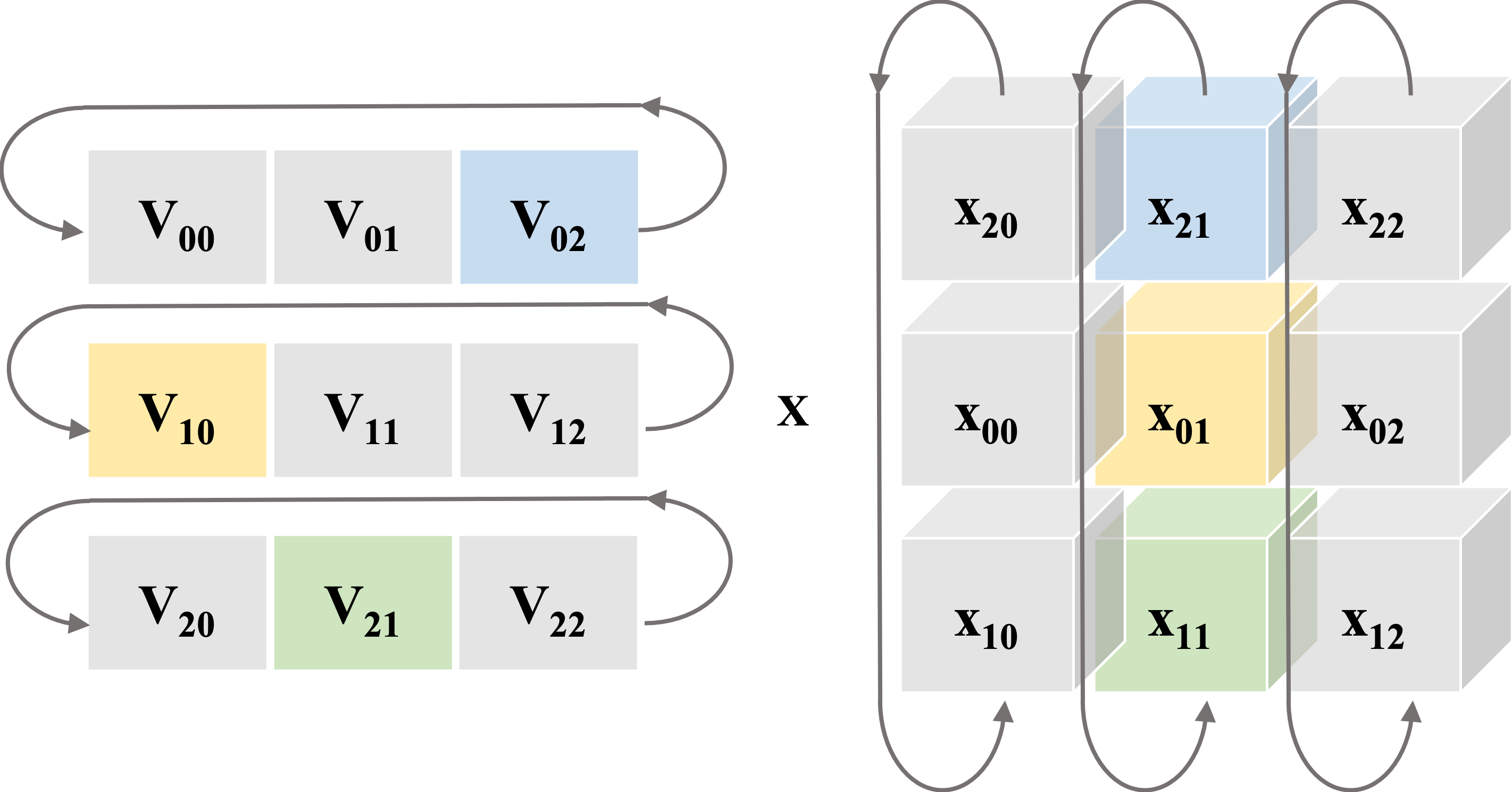}
		\caption{} 
	\end{subfigure}
	\caption{The one-shuffle scheme in the parallel algorithm based on KDFT is illustrated with a 3D example.
		We use $c_0$, $c_1$, and $c_2$ to denote three adjacent cores, the operations on which are highlighted with blue, yellow, and green, respectively. 
		The data decomposition results in the block of the input data $x_{01}$ and the slice of the Vandermonde matrix $\left[V_{00}, V_{01}, V_{02}\right]$ on core $c_0$, $x_{11}$ and $\left[V_{10}, V_{11}, V_{12}\right]$ on core $c_1$, 
		and $x_{21}$ and $\left[V_{20}, V_{21}, V_{22}\right]$ on core $c_2$.
		The steps involved in the one-shuffle scheme are: (a) computing $V_{00} \cdot x_{01}$ on core $c_0$,  $V_{11} \cdot x_{11}$ on core $c_1$, and $V_{22} \cdot x_{21}$ on core $c_2$ with
		$\cdot$ representing the operation of tensor contraction;
		(b) collectively permuting the inputs between two neighboring cores such that $x_{11}$ on core $c_0$, $x_{21}$ on core $c_1$, and $x_{01}$ on core $c_2$ and
		computing $V_{01} \cdot x_{11}$ on core $c_0$,  $V_{12} \cdot x_{21}$ on core $c_1$, and $V_{20} \cdot x_{01}$ on core $c_2$;
		(c) collectively permuting the inputs such that $x_{21}$ on core $c_0$, $x_{01}$ on core $c_1$, and $x_{11}$ on core $c_2$ and
		computing $V_{02} \cdot x_{21}$ on core $c_0$,  $V_{10} \cdot x_{01}$ on core $c_1$, and $V_{21} \cdot x_{11}$ on core $c_2$.
		The $\texttt{collective\_permute}$ operation in shuffling the input between neighboring TPU cores is with source-target pairs $(c_1,c_0)$, $(c_2,c_1)$, and $(c_0,c_2)$ in the form of $(\textrm{source},\textrm{target})$.}
	\label{one_shuffle}
\end{figure}

In this section, we provide details for implementing both KDFT and FFT on TPUs, including the data decomposition and the one-shuffle scheme.

\subsection{Data decomposition}
The data decomposition applied to the input data localizes the matrix multiplications on individual cores, which is critical to achieve high parallel efficiency on TPUs.
For a 3D DFT, the data decomposition is applied to the input data along all three dimensions.
The decomposition is based on a TPU computation shape $\left( P_1, P_2, P_3 \right)$ where $P_1$, $P_2$, and $P_3$ denote the number of TPU cores
along the first, the second, and the third dimension, respectively.
Given the TPU computation shape $\left( P_1, P_2, P_3 \right)$ and the input data of dimension $N_1 \times N_2 \times N_3$,
each TPU core contains a data block of dimension $\frac{N_1}{P_1} \times \frac{N_2}{P_2}  \times \frac{N_3}{P_3}$ as shown in Fig.\ \ref{load_balancing}(a).
The data decomposition is also applied to the Vandermonde matrix and is along the frequency domain.
As shown in Fig.\ \ref{load_balancing}(b), each core has a slice of the Vandermonde matrix with dimension $\frac{N_i}{P_i} \times N_i$, $i = 1, 2, 3$.
It is also shown in Fig.\ \ref{load_balancing} that each core is assigned an index $p$ along each dimension and $p_i = 0, 1, \cdots, P_i-1$,
where $i = 1, 2, 3$.
With the proposed data decomposition, the dense matrix multiplications of both KDFT and FFT are kept local within individual TPU cores and performed completely in parallel.

\subsection{One-shuffle scheme}
The one-shuffle scheme described in Algorithm \ref{one_shuffle_pseudocode} is used by both KDFT and FFT.
We use KDFT to illustrate the one-shuffle scheme.
There are two major operations in KDFT: the tensor contraction between the Vandermonde matrix and the input data; and the communication among TPU cores to send and receive the data.
The  tensor contraction is based on $\texttt{einsum}$ and the communication among TPU cores is with $\texttt{collective\_permute}$.
After one operation of tensor contraction, the block of the input data initially assigned on a TPU core is shuffled once with its neighboring core. 
The one-time shuffling takes place along the same direction on the interconnect network. 
As shown in Fig.\ \ref{load_balancing}(b), the DFT along the third dimension requires shuffling the blocks of the input data among the cores
that are grouped by the third dimension of the computation shape $P_3$.
In FFT, the one-shuffle scheme is used for applying the phase adjustment, in which
the Vandermonde matrix in Fig.\ \ref{load_balancing}(a) contains the
phase-shift information.

\begin{algorithm}
	\caption{The one-shuffle scheme}
	\label{one_shuffle_pseudocode}
	\begin{algorithmic}[1]
		\Function{$\textrm{one\_shuffle}$}{$\textrm{v}$, $\textrm{x}$, $\textrm{core\_idx}$, $\textrm{num\_cores}$, $\textrm{src\_tgt\_pairs}$}
		\State $ \textrm{iteration\_idx} \gets  0 $
		\State $ \textrm{slice\_idx} \gets \textrm{core\_idx} $
		\State $\textrm{x\_out} \gets \texttt{einsum} (\textrm{v}[\textrm{slice\_idx}], ~\textrm{x})$
		\State $\textrm{slice\_idx} \gets \texttt{mod}(\textrm{slice\_idx} + 1, ~\textrm{num\_cores})$
		\While {$ \textrm{iteration\_idx}  < \textrm{num\_cores} - 1$}
		\State $\textrm{x} \gets \texttt{collective\_permute} (\textrm{x}, ~\textrm{src\_tgt\_pairs})$
		\State $\textrm{x\_out} \gets \textrm{x\_out}~+~\texttt{einsum} (\textrm{v}[\textrm{slice\_idx}], ~\textrm{x})$
		\State $\textrm{slice\_idx} \gets \texttt{mod}(\textrm{slice\_idx} + 1, ~\textrm{num\_cores})$
		\State $ \textrm{iteration\_idx} \gets \textrm{iteration\_idx}  ~+~1 $
		\EndWhile \label{oneshufflewhile}
		\State  \textbf{return}  $\textrm{x\_out}$
		\EndFunction
	\end{algorithmic}
\end{algorithm}

With the one-shuffle scheme, sending and receiving data takes place simultaneously between two neighboring cores and along the same direction on the 2D toroidal network.
The one-shuffle scheme best utilizes the topology of the interconnect and hence its bandwidth.
Together with the high-speed and low-latency nature of the interconnects, the one-shuffle scheme requires minimal communication time.

\subsection{Implementation of the parallel algorithm based on KDFT}
Figure \ref{one_shuffle} illustrates the one-shuffle scheme in the parallel algorithm based on KDFT with a 3D example.
We use $c_0$, $c_1$, and $c_2$ to denote three adjacent TPU cores, the operations on which are highlighted with three different colors accordingly in Fig.\ \ref{one_shuffle}.
After the data decomposition, core $c_0$ contains a block of the input data $x_{01}$ and a slice of the Vandermonde matrix $\left[V_{00}, V_{01}, V_{02}\right]$,
core $c_1$ contains $x_{11}$ and $\left[V_{10}, V_{11}, V_{12}\right]$,
and core $c_2$ contains $x_{21}$ and $\left[V_{20}, V_{21}, V_{22}\right]$.
Note that the subscripts appearing in the block of the input data $x_{p_1, p_2, p_3}$ are core indices.
For simplicity, we ignore the core index on the third dimension, which is the same across cores $c_0$, $c_1$, and $c_2$.
With three $\texttt{einsum}$ and two $\texttt{collective\_permute}$ operations,
one obtains the partial DFT written as $V_{00} \cdot x_{01} + V_{01} \cdot x_{11} + V_{02} \cdot x_{21}$ on core $c_0$,
where $\cdot$ represents the tensor contraction.
The steps taken by the partial DFT computation along one dimension are as follows:
\begin{itemize}
	\item [1.] apply $\texttt{einsum}$ to compute $V_{00} \cdot x_{01}$ on core $c_0$,  $V_{11} \cdot x_{11}$ on core $c_1$, and $V_{22} \cdot x_{21}$ on core $c_2$ as shown in Fig.\ \ref{one_shuffle}(a);
	\item [2.] apply $\texttt{collective\_permute}$ to shuffle the input between neighboring TPU cores with source-target pairs $(c_1,c_0)$, $(c_2,c_1)$, and $(c_0,c_2)$ in the form of $(\textrm{source},\textrm{target})$
	such that core $c_0$ contains $x_{11}$, core $c_1$ contains $x_{21}$, and core $c_2$ contains $x_{01}$ as shown in Fig.\ \ref{one_shuffle}(b);
	\item [3.] apply $\texttt{einsum}$ to compute $V_{01} \cdot x_{11}$ on core $c_0$,  $V_{12} \cdot x_{21}$ on core $c_1$, and $V_{20} \cdot x_{01}$ on core $c_2$ and add the results from step 1;
	\item [4.] apply $\texttt{collective\_permute}$ with source-target pairs $(c_1,c_0)$, $(c_2,c_1)$, $(c_0,c_2)$, after which core $c_0$ contains $x_{21}$, core $c_1$ contains $x_{01}$, and core $c_2$ contains $x_{11}$ as shown in Fig.\ \ref{one_shuffle}(c);
	\item [5.] apply $\texttt{einsum}$ to compute $V_{02} \cdot x_{21}$ on core $c_0$,  $V_{10} \cdot x_{01}$ on core $c_1$, and $V_{21} \cdot x_{11}$ on core $c_2$ and add the results from step 3.
\end{itemize}

\subsection{Implementation of the parallel algorithm based on FFT}
\begin{figure}[]
	\centering
	\begin{subfigure}{\linewidth}
		\centering
		\includegraphics[width=\linewidth]{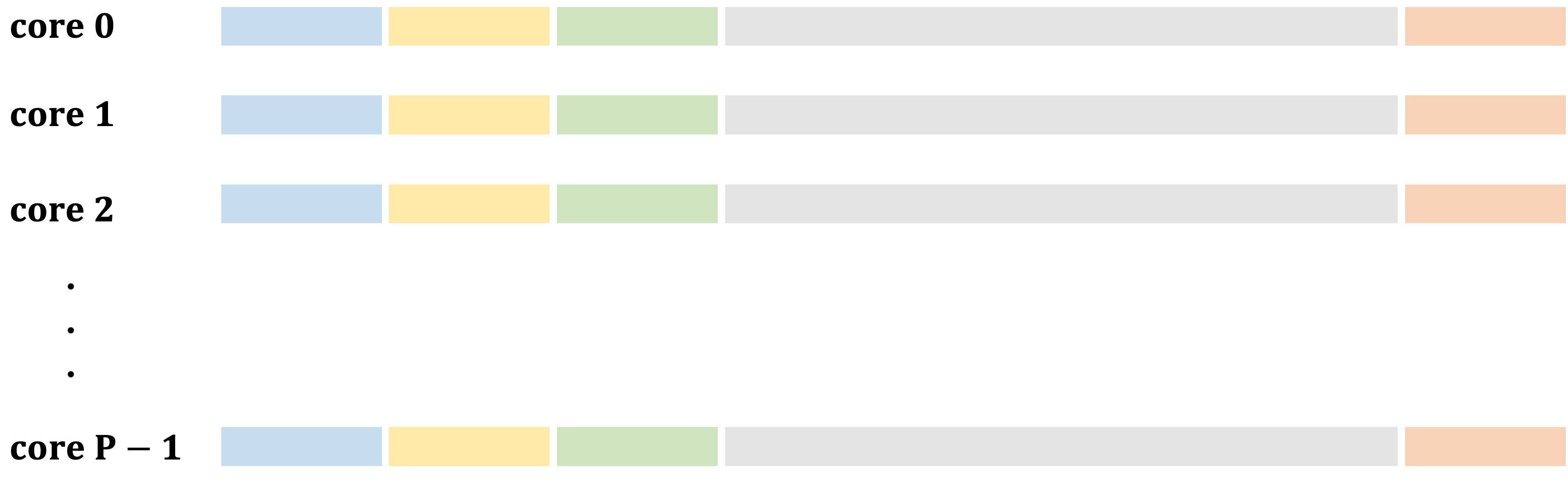}
		\caption{} 
	\end{subfigure} \hfill
	\begin{subfigure}{\linewidth}
		\centering
		\includegraphics[width=\linewidth]{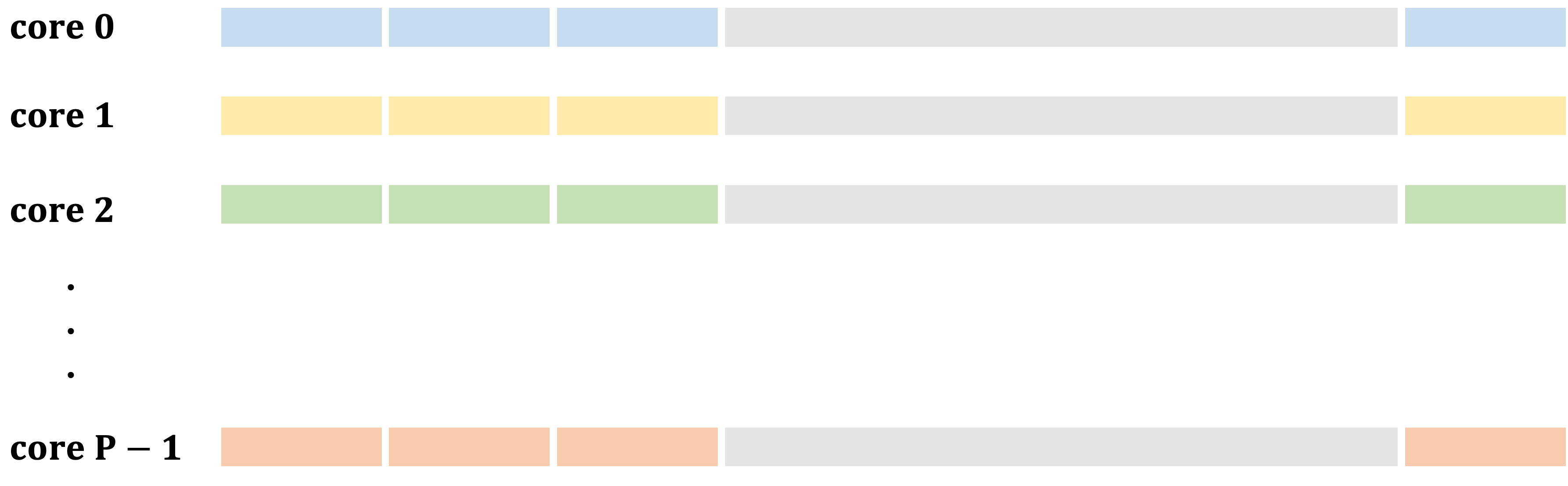}
		\caption{} 
	\end{subfigure}
	\begin{subfigure}{\linewidth}
		\centering
		\includegraphics[width=\linewidth]{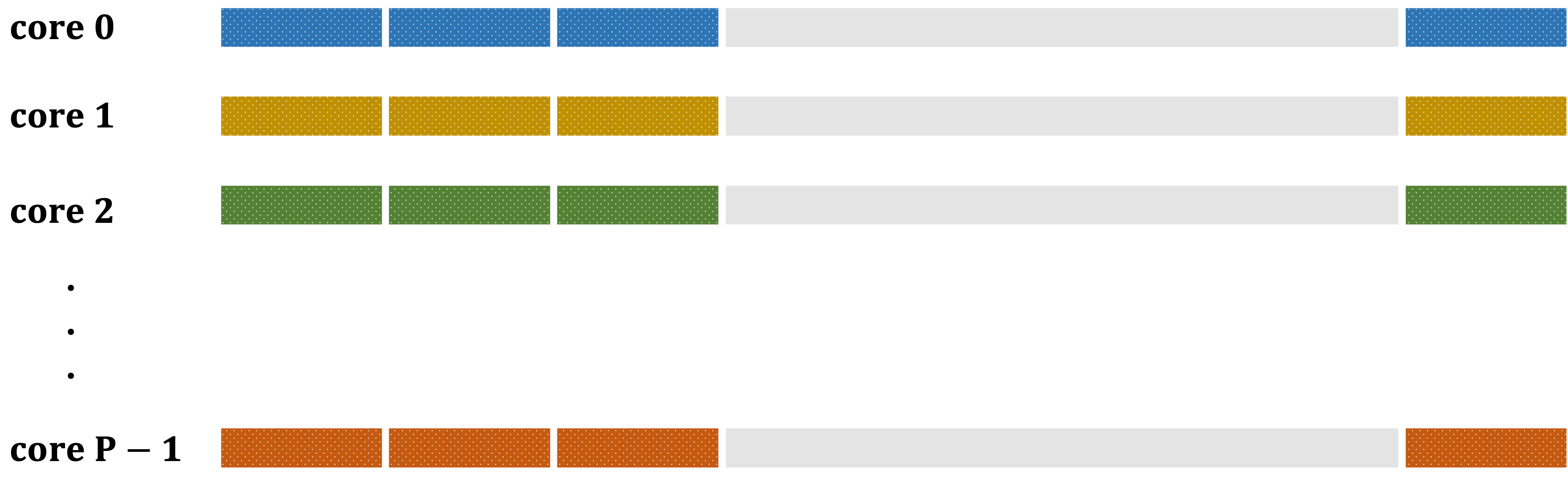}
		\caption{} 
	\end{subfigure}
	\begin{subfigure}{\linewidth}
		\centering
		\includegraphics[width=\linewidth]{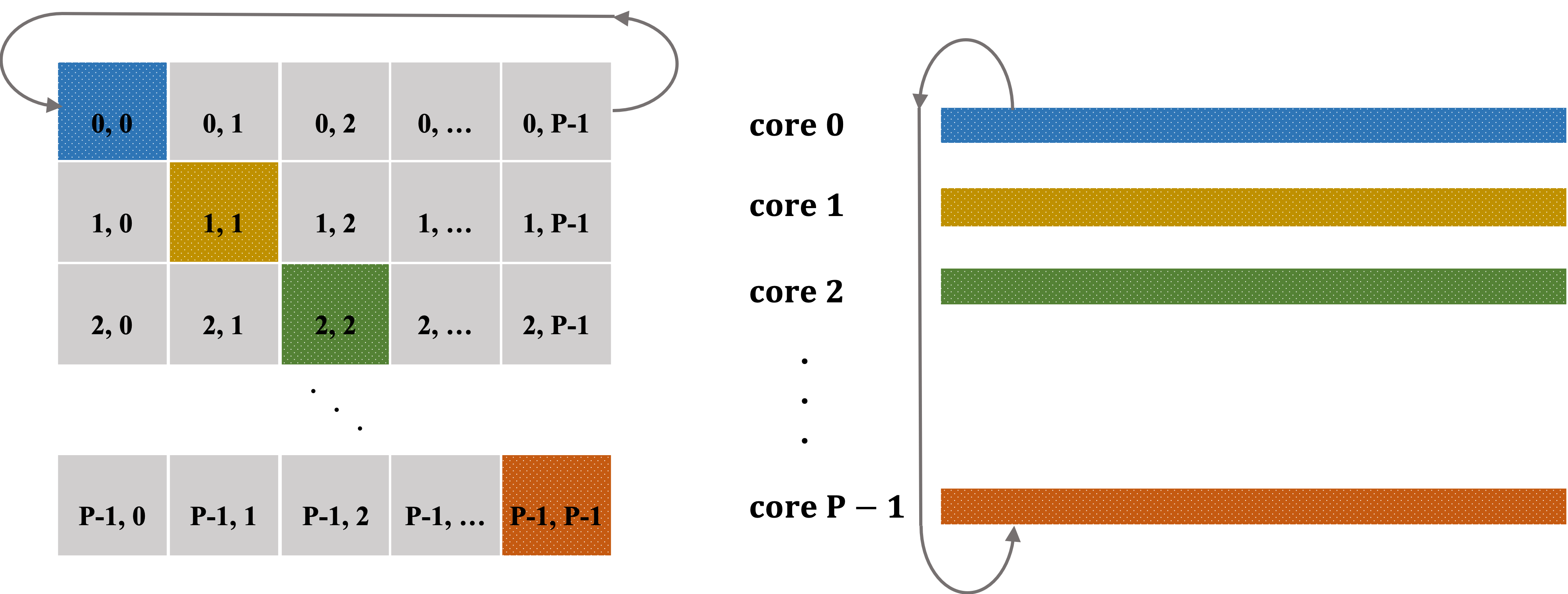}
		\caption{} 
	\end{subfigure}
	\caption{Four steps for a 1D FFT on TPUs: (a) the data decomposition, (b) the gathering of the input,
		(c) the transform performed locally on individual cores, and (d) the phase adjustment through the one-shuffle scheme. 
		The pair of the indices in (d) represents the source and target pairs used by $\texttt{collective\_permute}$.}
	\label{parallel_fft}
\end{figure}

Figure \ref{parallel_fft} illustrates the four steps of a 1D FFT on TPUs: the data decomposition, the gathering of the input,
the local transform, and the phase adjustment.
Figure \ref{parallel_fft}(a) shows the input assigned to individual cores after the data decomposition.
In order to achieve an in-order transform, to be specific, the ordering of the obtained results  in the transform domain remains the same as that in the input,
it requires a local re-ordering of the input prior to the transform, which can be achieved through $\texttt{einsum}$.
The re-ordering operation is local within individual TPU cores.
The gathering of the input as shown in Fig.\ \ref{parallel_fft}(b) is implemented with $\texttt{all\_to\_all}$.
After the gathering, the Cooley-Tukey-algorithm-based transform is performed locally on individual cores, which is implemented with $\texttt{tf.signal.fft}$ as shown in Fig.\ \ref{parallel_fft}(c).
At last, the locally-obtained transform results are summed over all the cores with phase adjustment, which is achieved through the one-shuffle scheme.
It can be seen that in the 1D FFT on TPUs, the communication is required in
gathering the input and applying the phase adjustment.
Higher dimensional FFT such as 2D and 3D can be achieved by repeating the 1D FFT along the corresponding dimensions.

\section{Parallel Efficiency Analysis}
\begin{figure*}[t!]
	\includegraphics[width=14cm, height=5cm]{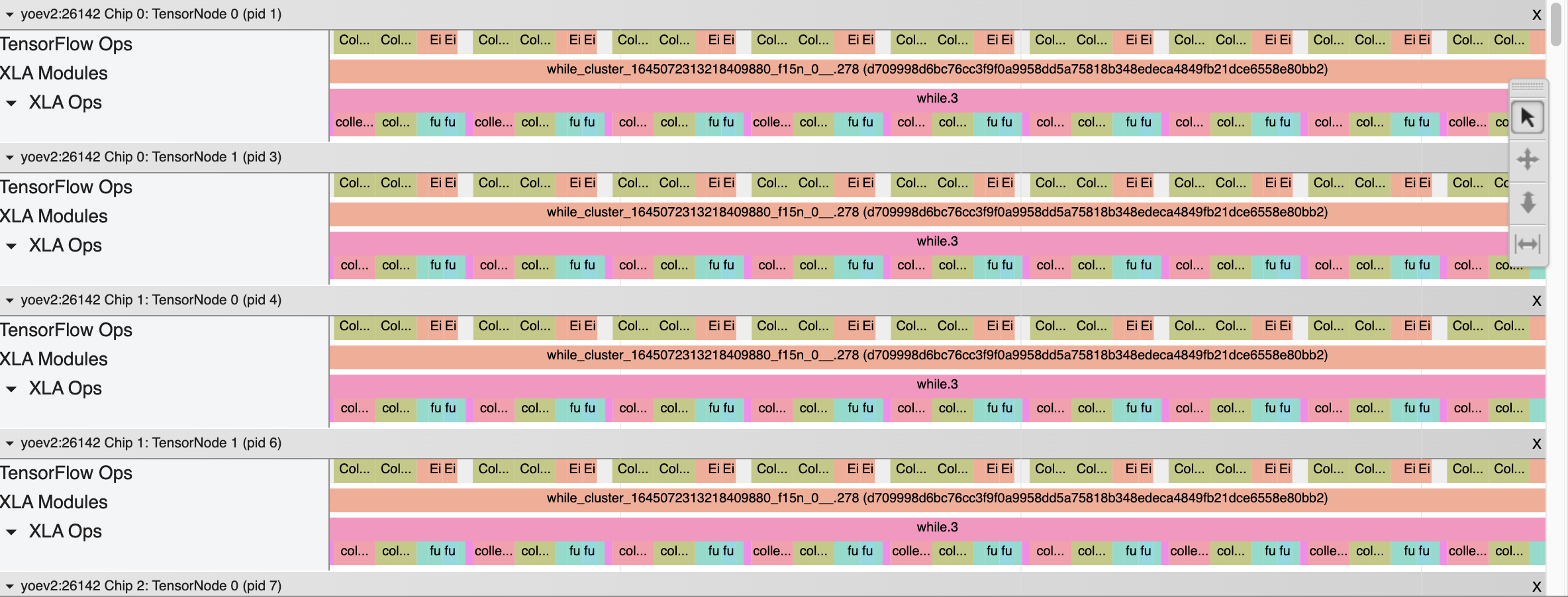}
	\centering
	\caption{A screenshot of the trace viewer from the TPU profiling tool.}
	\label{xprof}
\end{figure*}

In this section, both the strong and weak scaling analyses are provided to demonstrate the efficiency of the proposed two DFT parallel algorithms on TPUs.
For the strong scaling analysis, the problem size is kept the same as proportionally more TPU cores are employed.
For the weak scaling analysis, the number of TPU cores remains the same as the problem size increases.
The TPU profiling tool \cite{tpuv3}, which provides information on the utilization of the hardware and
the efficiency of individual operations at the program level is used to analyze the performance of DFT on TPUs.
A screenshot of the trace viewer from the TPU profiling tool is shown in Fig.\ \ref{xprof}.
With the profiling tool, one can breakdown the operations at the HLO level, which is quite helpful in identifying the bottleneck of the
parallel efficiency and making improvements to the algorithm designs.

\subsection{Strong scaling analysis of 2D KDFT}
\begin{figure}[t!]
	\includegraphics[width=8cm, height=6cm]{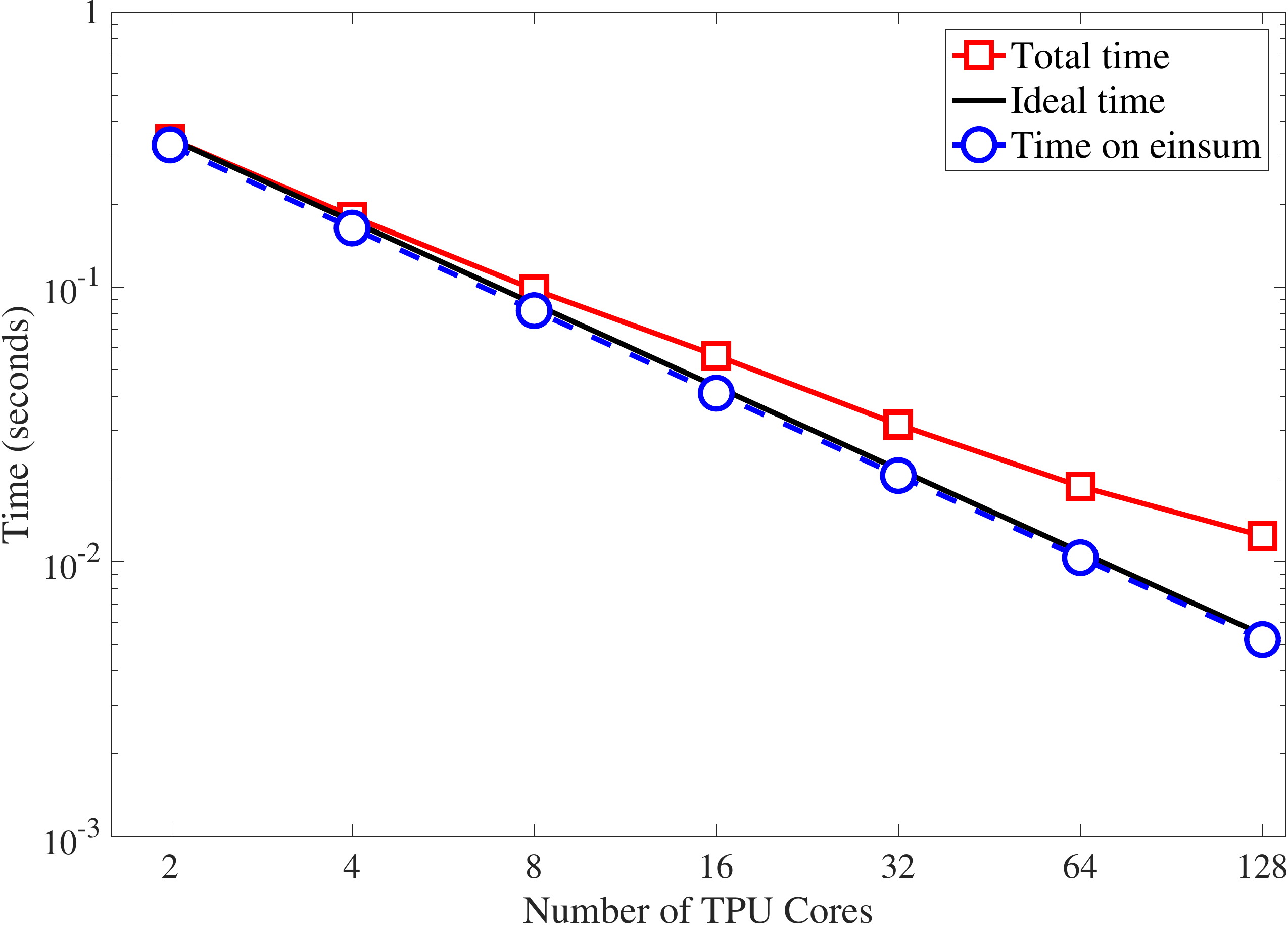}
	\centering
	\caption{ The computation time of the 2D KDFT with up to 128 TPU cores on an example of dimension $8192 \times 8192$.}
	\label{strongscaling_dft2d}
\end{figure}

Figure \ref{strongscaling_dft2d} shows the computation time of the 2D KDFT with up to 128 TPU cores on an example of dimension $8192 \times 8192$.
It can be seen from Fig.\ \ref{strongscaling_dft2d} that a close-to-linear scaling of the computation time with respect to the number of TPU cores is achieved.
As a reference, the ideal computation time from the linear scaling is provided in Fig.\ \ref{strongscaling_dft2d}, which is defined by
\begin{equation}
\textrm{ideal time} = \frac{T_2}{\frac{N_{\textrm{core}}}{2}},
\end{equation}
where $T_2$ denotes the total computation time with two TPU cores and ${N_{\textrm{core}}}$ is the total number of TPU cores being used.
As mentioned in the parallel implementation, the total computation time consists of two parts: the time of matrix multiplications, or $\texttt{einsum}$ to be specific,
and the communication time of sending and receiving the block of input data across TPU cores.
It can be seen from Fig.\ \ref{strongscaling_dft2d} that the time of matrix multiplications scales linearly with respect to the total number of TPU cores.
This is because the matrix multiplications are kept completely local within individual cores.
The computation time of the 2D KDFT scales approximately linearly with respect to the number of TPU cores,
with the gap between the actual and the ideal computation time arising from the communication among TPU cores.

\subsection{Strong scaling analysis of 3D KDFT}
\begin{figure}[t!]
	\includegraphics[width=8cm, height=6cm]{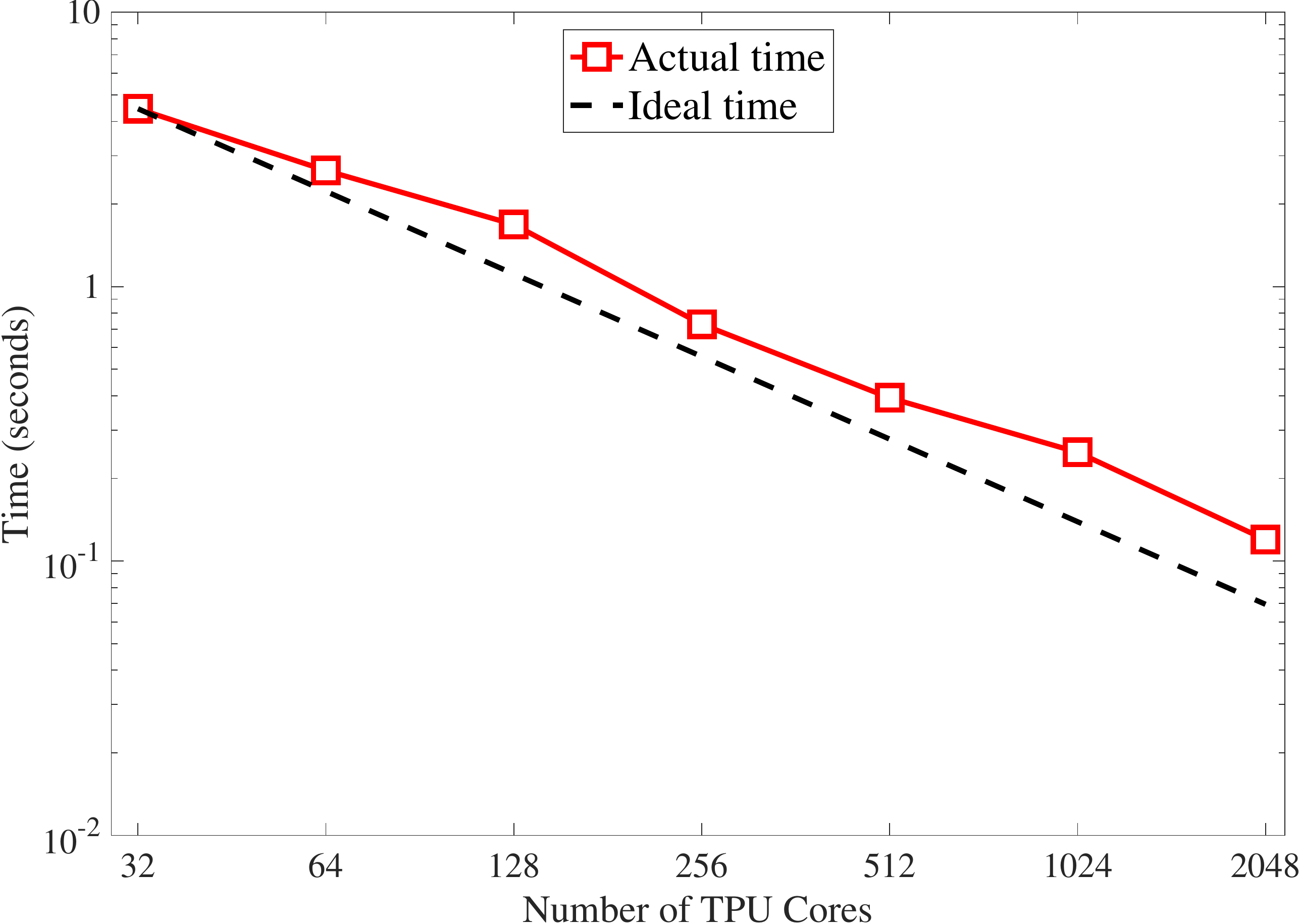}
	\centering
	\caption{The computation time of the 3D KDFT with up to 2048 TPU cores on an example of dimension $2048 \times 2048 \times 2048$.}
	\label{strongscaling_dft3d}
\end{figure}
The parallel efficiency of the 3D KDFT is demonstrated through an example of dimension $2048 \times 2048 \times 2048$.
Similar to the 2D case, the problem size is also fixed as proportionally more TPU cores are employed.
The total computation time is depicted in Fig.\ \ref{strongscaling_dft3d}.
As a reference, the ideal computation time from linear scaling is provided in Fig.\ \ref{strongscaling_dft3d}, which is defined by
\begin{equation}
\textrm{ideal time} = \frac{T_{32}}{\frac{N_{\textrm{core}}}{32}},
\end{equation}
where $T_{32}$ denotes the total computation time with 32 TPU cores.
It can be seen from Fig.\ \ref{strongscaling_dft3d} that the computation time scales approximately linearly with respect to the number of TPU cores.

\begin{figure}[t!]
	\includegraphics[width=8cm, height=6cm]{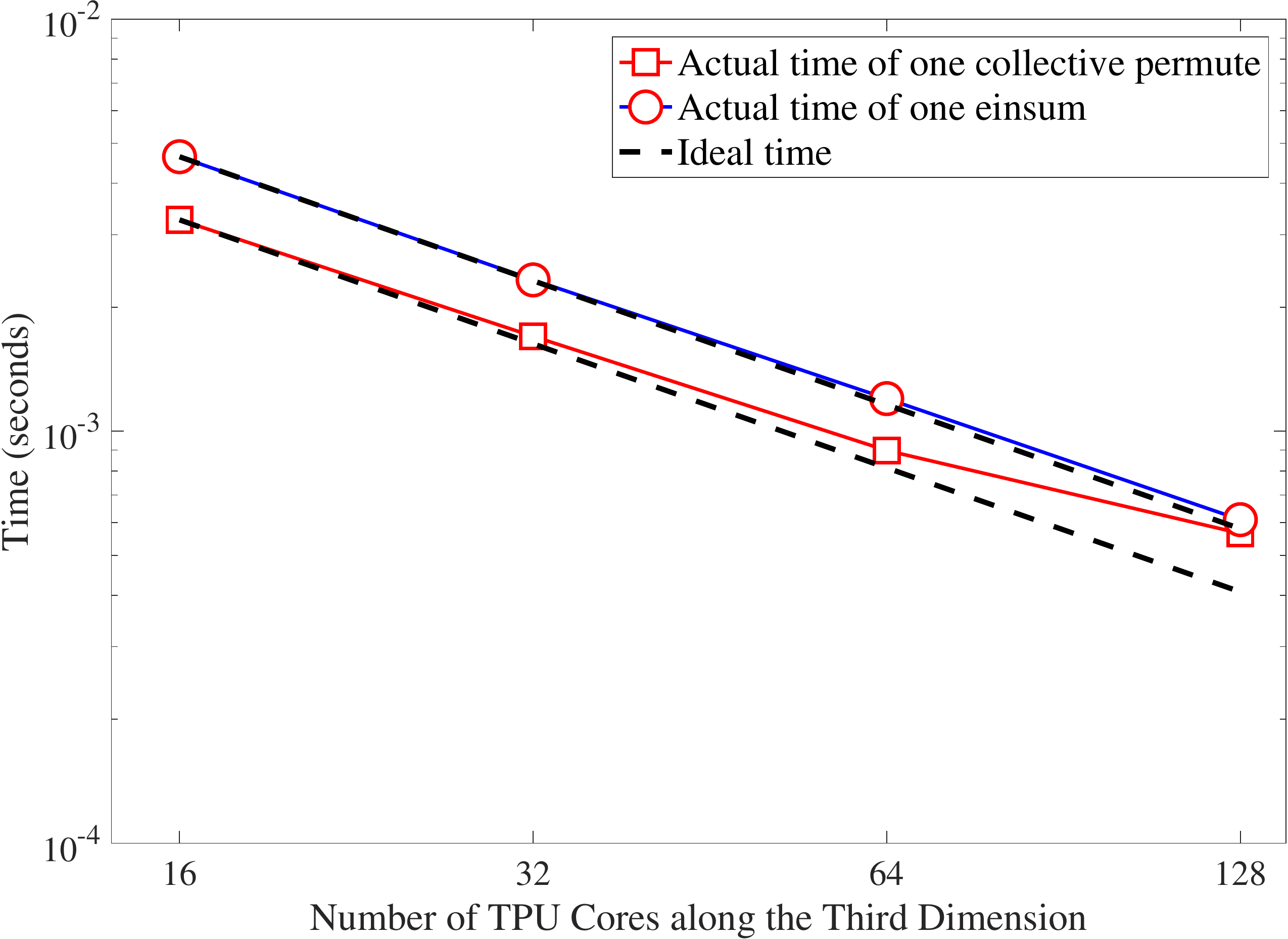}
	\centering
	\caption{The computation time of one single operation of $\texttt{collective\_permute}$ and $\texttt{einsum}$, respectively, in the 3D KDFT along the third dimension with respect to the number of TPU cores. }
	\label{strongscaling_dft3d_step}
\end{figure}

The gap between the actual and the ideal computation time in the 3D case also results from the communication among TPU cores.
As mentioned in the parallel implementation, the data decomposition is applied to the input data along all the three dimensions with a TPU computation shape.
The computation shape in this example has the form of $\left(4, 4, n_2\right)$ with four TPU cores along the first dimension, four along the second dimension, and $n_2$ along the third dimension.
It is indeed the number of TPU cores along the third dimension that varies in Fig.\ \ref{strongscaling_dft3d}.
For example, the computation shapes $\left(4, 4, 16\right)$ and $\left(4, 4, 32 \right)$ are corresponding to  256 and 512 TPU cores, respectively.
As the number of TPU cores along the third dimension doubles itself, the size of the input data contained on each core is reduced by half.
As a result, the computation time associated with a single operation of $\texttt{collective\_permute}$ or $\texttt{einsum}$ is also reduced by half, which is shown in Fig.\ \ref{strongscaling_dft3d_step}.
However, as more cores are being used, the total number of $\texttt{collective\_permute}$ operations increases.
For example,  it requires a total number of 31 $\texttt{collective\_permute}$ operations in the Fourier transform along the third dimension in the case of 512 TPU cores or with the TPU computation shape $\left(4, 4, 32\right)$, whereas only 15 $\texttt{collective\_permute}$ operations are required in the case of 256 TPU cores or with the TPU computation shape $\left(4, 4, 16\right)$.
It can be seen that even though the time associated with one single $\texttt{collective\_permute}$ operation decreases when more TPU cores are used,
the total communication time for the DFT along the third dimension does not change. 
This explains the gap between the total and the ideal computation time in Fig.\ \ref{strongscaling_dft3d}.

\subsection{Strong scaling analysis of 3D FFT}
\begin{figure}[t!]
	\includegraphics[width=8cm, height=6cm]{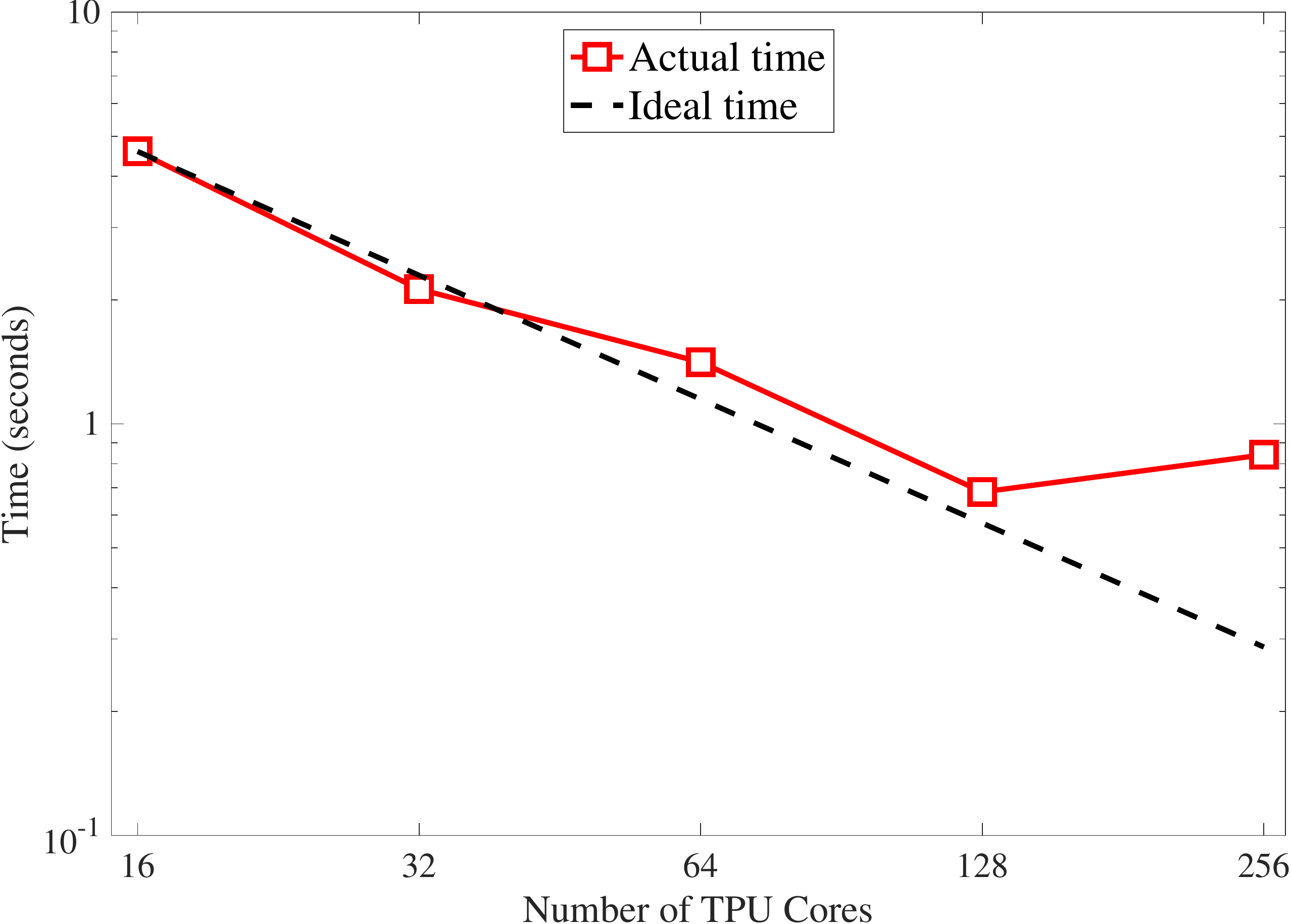}
	\centering
	\caption{The computation time of the 3D FFT with up to 256 TPU cores on an example of dimension $2048 \times 2048 \times 2048$.}
	\label{strongscaling_fft3d}
\end{figure}

\begin{figure}[t!]
	\includegraphics[width=8cm, height=4.6cm]{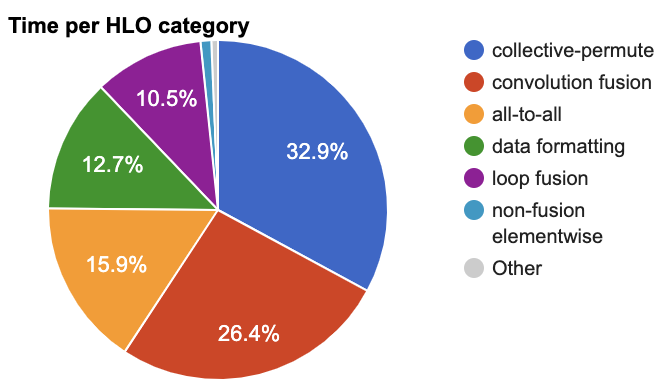}
	\centering
	\caption{The breakdown of the computation time with the TPU profiling tool for the 3D FFT with 256 TPU cores on an example of dimension $2048 \times 2048 \times 2048$.}
	\label{hlo_stats_fft3d}
\end{figure}

The parallel efficiency of the 3D FFT is demonstrated through an example of dimension $2048 \times 2048 \times 2048$.
The problem size is  fixed as proportionally more TPU cores are employed.
The total computation time is depicted in Fig.\ \ref{strongscaling_fft3d}.
As a reference, the ideal computation time from linear scaling is provided in Fig.\ \ref{strongscaling_fft3d}, which is defined by
\begin{equation}
	\textrm{ideal time} = \frac{T_{16}}{\frac{N_{\textrm{core}}}{16}},
\end{equation}
where $T_{16}$ denotes the total computation time with 16 TPU cores.
As shown in Fig.\ \ref{strongscaling_fft3d} that close-to-linear scaling is observed in the 3D FFT computation on TPUs up to 128 cores.
The example of size $2048 \times 2048 \times 2048$ is considered small for more than 128 cores.
The breakdown of the total computation time from the TPU profiling tool is provided in Fig.\ \ref{hlo_stats_fft3d}.
It can be seen that the communication time consisting of both $\texttt{all\_to\_all}$ and $\texttt{collective\_permute}$ starts dominating the total computation time.

\begin{table}[]
	\normalsize
	\renewcommand{\arraystretch}{1.8}
	\centering
	\caption{Computation time of the 3D KDFT and FFT on a full TPU Pod with 2048 TPU cores.}
	\label{3ddft_fft_full_pod}
	\begin{tabular}{|c|c|c|c|}
		\hline
		\multirow{2}{*}{No.} & \multirow{2}{*}{Problem size} & \multicolumn{2}{l|}{Time (seconds)} \\ \cline{3-4} 
		&                               & KDFT               & FFT             \\ \hline
		1                 & $8192 \times 8192 \times 8192$                      & 12.66             & 8.30                \\ \hline
		2                 & $4096 \times 4096 \times 4096$                      & 1.07              & 1.01             \\ \hline
		3                 & $2048 \times 2048 \times 2048$                      & 0.120              &  0.118             \\ \hline
		4                 & $1024 \times 1024 \times 1024$                      & 0.0220              & 0.0148             \\ \hline
	\end{tabular}
\end{table}

\subsection{3D KDFT and FFT on a Full TPU Pod}
In addition to the strong scaling analysis, the computation time of a few 3D DFT and FFT examples on a full TPU Pod with 2048 cores is provided in Table \ref{3ddft_fft_full_pod}.
As the problem size increases from $2048 \times 2048 \times 2048$ to $4096 \times 4096 \times 4096$, the computation time of DFT increases 9.7 times.
Similarly, the computation time of DFT increases 11.8 times when the problem size increases from $4096 \times 4096 \times 4096$ to $8192 \times 8192 \times 8192$.
As a comparison, the computation time of FFT scales up approximately eight times as the problem size increases by eight times.
For the two problems of sizes $2048 \times 2048 \times 2048$ and $4096 \times 4096 \times 4096$, the total computation time of KDFT is about the same as that of FFT.
Given the computation complexity difference between KFDFT and FFT,
it demonstrates TPU's strength in matrix multiplications.

\section{Conclusion and Discussion}
In this work, we proposed and implemented two parallel algorithms of DFT on TPUs, to be specific, KDFT and FFT.
The formulation of KDFT is based on the Kronecker product.
The formulation of FFT is based on the Cooley-Tukey algorithm and the phase adjustment.
Both formulations take full advantage of TPU's strength in matrix multiplications.
The implementation is in TensorFlow.
Through implementing the two parallel algorithms, TPU---the domain-specific hardware accelerator for machine learning applications---is employed in the parallel computing of large-scale DFT.
The data decomposition adopted in both parallel algorithms enables the localization of
the dense matrix multiplications on individual TPU cores,  which can be performed completely in parallel.
As for the communication among TPU cores, the one-shuffle scheme is designed based on the TPU interconnect topology,
with which sending and receiving data takes place simultaneously between two neighboring cores and along the same direction on the interconnect network.
The one-shuffle scheme requires minimal communication time among TPU cores and achieves very high parallel efficiency.
Scaling analysis is provided to understand the parallel efficiency of the proposed DFT algorithms on TPUs.

With the demonstrated computation efficiency, the large-scale DFT on TPUs opens an array of opportunities for scientific computing. 
One possible future work is to integrate the DFT on TPUs with medical image reconstruction,
where nonuniform Fourier transform is extensively used.
Another future work is to extend the two proposed algorithms into a framework and address large-scale DFT of higher dimensions.
Finally, the precision of matrix multiplications in this work can be improved from float32 to float64.

\section{Acknowledgment}
We would like to thank David Majnemer, Reid Tatge, Dehao Chen, Yusef Shafi, Damien Pierce, James Lottes, Matthias Ihme, and Rene Salmon for valuable
discussions and helpful comments, which have greatly improved the paper.

\bibliographystyle{IEEEtran}
\bibliography{bibliography}

\end{document}